# Nonvolatile bipolar resistive switching in Au/BiFeO$_3$/Pt

Yao Shuai,[1] Shengqiang Zhou,[1,2] Danilo Bürger,[1] Manfred Helm,[1] and Heidemarie Schmidt[1]

[1]*Institute of Ion Beam Physics and Materials Research, Helmholtz-Zentrum Dresden-Rossendorf (HZDR), P. O. Box 510119, Dresden 01314, Germany*

[2]*State Key Laboratory of Nuclear Physics and Technology, School of Physics, Peking University, Beijing 100871, China*

**Abstract:** Nonvolatile bipolar resistive switching has been observed in an Au/BiFeO$_3$/Pt structure, where a Schottky contact and a quasi-Ohmic contact were formed at the Au/BiFeO$_3$ and BiFeO$_3$/Pt interface, respectively. By changing the polarity of the external voltage, the Au/BiFeO$_3$/Pt is switched between two stable resistance states without an electroforming process. The resistance ratio is larger than two orders of magnitude. The resistive switching is understood by the electric field – induced carriers trapping and detrapping, which changes the depletion layer thickness at the Au/BiFeO$_3$ interface.

I. INTRODUCTION

Resistive switching in oxides has attracted increasing attention due to the potential application for nonvolatile memory devices.[1-5] Although different resistive switching mechanisms have been reported in literature, the underlying physical mechanism is still a controversial issue. BiFeO$_3$ (BFO) is currently the most investigated multiferroic material which shows magnetoelectric coupling effects.[6-9] There have been many reports on the multiferroic properties of BFO, however, only a few reports focused on its resistive switching behavior.[10-12]

Resistive switching has been reported to be due to thermal,[1] ionic,[2,3] or electronic effects.[4,5] In an asymmetric metal-insulator-metal structure where a Schottky contact and an Ohmic contact are formed at the two interfaces, respectively. It is generally believed that the Schottky interface dominates the bipolar resistive switching behavior.[4,5,12] However, the electrical characterization which can reveal the dynamic change of the depletion layer has rarely been reported, and the mechanism of the switching nonvolatility in oxides is unclear. Besides, little attention has been paid to the voltage polarity for the high resistance state (HRS) and low resistance state (LRS).[13] For a "counter eightwise" I-V hysteresis, the reverse bias of the Schottky contact attracts the oxygen vacancies (OVs) towards the anode, which narrows the depletion region or forms local conductive filaments and turns the structure into LRS.[14,15] However, it is much more difficult to explain the "eightwise" I-V curve, where the reverse bias induces HRS instead of LRS.[13]

In the present work, a bipolar resistive switching showing "eightwise" I-V hysteresis has been observed in an Au/BFO/Pt stack without any electroforming step. The nonvolatility of the switching is demonstrated by resistance measurements after writing pulses. Based on the analysis of the transport characteristics during the switching, we discuss the switching mechanism and the physical origin of nonvolatility in Au/BFO/Pt structures.

## II. EXPERIMENT

BFO thin films were grown on Pt/Ti/SiO$_2$/Si substrates by pulsed laser deposition (PLD). The detailed PLD parameters are summarized in Tab. 1. The crystalline phase was checked by XRD and no impurity phases have been observed within the XRD detection limit. Au top contacts with an area of 0.1 mm$^2$ were fabricated by RF sputtering at room temperature using a metal shadow mask. The current-voltage (I-V) and pulsed voltage measurements were carried out using a Keithley 2400. The capacitance-voltage (C-V) was measured by an impedance analyzer (Agilent 4294A).

## III. RESULTS AND DISCUSSION

As illustrated in Fig. 1, the I-V curve shows rectifying characteristics. Here the forward bias is defined as a positive bias applied to the Au top contact. The sequence of the voltage is: 0V → +11V → -11V → 0V. The current shows a strong asymmetric shape with a rectification factor of 1200 at ±11 V. Considering that donor-like OVs are always present in PLD-grown BFO thin films, the BFO thin film is *n*-type conducting.[16] Combining with voltage polarity of the I-V curves, we conclude that a Schottky contact has been formed at the Au/BFO interface, while the BFO/Pt is an Ohmic-like contact.

Beside the rectifying behavior, also bipolar resistive switching is observed (Fig. 1). An "eightwise" I-V hysteresis can be observed. The structure shows homogeneous performance, which is revealed by measuring different cells on the same wafer. No electroforming process is necessary before the observation of the hysteretic I-V curves. The pristine sample always exhibits high resistance state (HRS), by applying a positive voltage, the stack can be switched to low resistance state (LRS), while the HRS is reset by a negative voltage. The current ratio between the LRS and HRS is more than two orders of magnitude at a voltage of +2 V.

To check the nonvolatility of the resistive switching observed in the I-V curves, pulsed voltage measurements were carried out. Writing pulses of ±11 V were applied to switch the resistance state of the stack between HRS and LRS, while the resistance state of the stack was detected using a small pulse voltage of +2V (reading pulse) (Fig. 2). The reading pulse was repeated for one thousand times to read the resistance of both HRS and LRS after either of them was written with -11 V and +11 V, respectively. In the as switched case, the LRS exhibits an increasing resistance, while the HRS is stable. However, in the storage retention testing, which was carried out by reading the resistance state 24 and 72 hours after either state was induced, the HRS is always stable, while the LRS becomes stable after 72 hours. Additionally, it is observed that after 72 hours the resistance ratio $R_{HRS}/R_{LRS}$ decreased from 631 to 336 (inset of Fig. 2). However, both states exhibit no significant resistance change if the retention time is further increased. The retention measurement indicates that both resistance states are stable enough in time for memory applications. Note that in the pulsed voltage measurement, although a +2 V reading pulse has been repeatedly applied for resistance detection, this relatively small voltage does not vary the resistance of the stack for both the HRS and LRS. That is important for a practical memory application which requires a stable resistance state even if the unit is read for many times.

To elucidate the switching mechanism, it is necessary to investigate the conduction mechanism for both the HRS and LRS. As shown in Fig. 3(a) and (b), the I-V curves have been replotted on a $\log(J/E) \sim E^{1/2}$, $\log(J/E^2) \sim 1/E$, and $\log(J) \sim E^{1/2}$ scale, representing the Poole-Frenkel emission (PF), Fowler-Nordheim (FN), and Schottky emission (SE), respectively. The optical dielectric constant K was extracted from fitting the slope using Eqs. (1-2) which describe the SE and PF conduction mechanism, respectively:[17]

$$J = A^* T^2 \exp-(\frac{\varphi_b}{kT} - \frac{q}{kT}\sqrt{\frac{qE}{4\pi\varepsilon_0 K}}) \quad \text{SE} \quad (1)$$

$$J = BE \exp-(\frac{E_I}{kT} - \frac{q}{kT}\sqrt{\frac{qE}{\pi\varepsilon_0 K}}) \quad \text{PF} \quad (2)$$

The index of reflection n of BFO thin films has been reported to be 2.5,[18] thus an optical dielectric constant value $K=n^2=6.25$ is expected. By comparing the fitted and expected K values, the conduction mechanism of the BFO thin film in the four different voltage ranges, labeled (1) to (4) in Fig. 1, can be deduced. In the voltage range (1) from 0 V to +11 V, the BFO thin film reveals PF conduction as the extracted optical dielectric constant K [Fig. 3(a)] is close to 6.25. Note that FN conduction is present at higher electric fields, as indicated by the negative linear fitting in the inset of Fig. 3(a). In the voltage range (2) from +11 V to 0 V, the BFO thin film exhibits FN conduction, which is suggested by the negative linear fitting as shown in the inset of Fig. 3(a). In the negative sweeping down (voltage range (3)) and sweeping up (voltage range (4)) range, the BFO thin film reveals SE conduction, as the slopes agree with the value of 6.25 which corresponds to the optical dielectric constant of BFO [Fig. 3(b)].

In BFO thin films the Poole-Frenkel ($J_{PF}$) and Fowler-Nordheim ($J_{FN}$) conduction mechanism exist in series and the dominant conduction mechanism is the one which has the larger resistance.[19] As discussed above, the dominant conduction is PF in HRS and FN in LRS, respectively. Therefore, $J_{PF}$ is lower in HRS than in LRS, revealing that electron hopping is more difficult in HRS than in LRS. It is well known that the electron hopping in BFO occurs between $Fe^{2+}$ and $Fe^{3+}$, thus more $Fe^{2+}$ ions can cause a larger $J_{PF}$. Note that the presence of $Fe^{2+}$ is also a prerequisite for charge compensation of the OVs, i.e. more $Fe^{2+}$ ions also imply more OVs.[19] When a positively charged OV attracts an electron, the $Fe^{3+}$ in the vicinity of this OV is reduced to $Fe^{2+}$.[20] Therefore, the OVs act as electron trapping centers in the BFO thin film. In LRS, more $Fe^{2+}$ are present, which means more trapping centers are occupied by electrons, favoring electron hopping. Only the Schottky interface limits the current.

Due to the Ohmic character of the bottom contact, the electrons are easily injected into the depletion layer at the Au/BFO interface from the Pt, significantly lowering the Schottky barrier. Therefore, tunneling dominates when the depletion region is thin enough [inset of Fig. 3(a)]. When reversely biasing the Schottky contact, the electrons are blocked, such that much fewer electrons can be injected into the BFO thin film from the Au top electrode. The depletion region extends as the reverse bias increases. The trapped electrons inside the BFO thin film are released and emitted into the bottom Pt electrode, and iron ions are reduced from $Fe^{2+}$ to $Fe^{3+}$ which destroys the electron hopping paths. The stack is therefore turned to HRS.

This trap charging and decharging has been used to explain the resistive switching behavior in Ti/$Sm_{0.7}Ca_{0.3}MnO_3$[4], $SrRuO_3$/$SrTi_{0.99}Nb_{0.01}O_3$[5], and Ti/$Pr_{0.7}Ca_{0.3}MnO_3$[21]. Note that although these junctions are consisting of different types of material, the polarities for HRS and LRS are the same, *i. e.* forward (reverse) bias of the Schottky contact induces LRS (HRS), which produces an "eightwise" I-V curve. We stress here that in many other reports where a "counter-eightwise" I-V curve has been observed, different mechanisms are proposed, for example, the migration of oxygen vacancies.[13-15] Therefore, the polarity for the HRS and LRS needs to be taken into account carefully when investigating the switching mechanism.

It should be emphasized that the asymmetric contact barrier, *i.e.*, the Schottky and quasi-Ohmic contact at the Au/BFO and BFO/Pt interface, respectively, is the necessary condition for the above discussed trapping and detrapping process, which leads to the resistive switching. The I-V curve between two neighboring Au contacts (symmetric contact barriers), where the detrapping and trapping processes also happen upon applying a bias, reveals no hysteresis (not shown). However, in the asymmetric contact geometry, the top Schottky contact blocks the electron injection under reverse bias, while the bottom Ohmic-like interface is transparent for electrons under forward bias. Therefore, the amount of injected electrons varies when switching the voltage polarity, which consequently changes the depletion thickness.

Although the Schottky interface-related switching has been reported before, a direct evidence which shows a clear relationship between the I-V and the interface effect is still lacking. To get further insight into the interface effect on the hysteretic I-V curve, the change of the depletion width is demonstrated by the capacitance-voltage (C-V) data recorded at 1 kHz (Fig. 4) with an ac test level of 50 mV. The C-V curve shows an obvious hysteresis behavior, revealing different depletion thicknesses for LRS and HRS at the same applied bias.[22] The abrupt drop of the capacitance at the high positive voltage indicates the collapse of the depletion region. Note that the capacitance peak at +0.9 V indicates a large change of charges with the ac test level when the dc voltage sweeps down. Further investigation is needed for this abnormal peak.

The nonvolatility of the switching can be explained by means of a dynamic equilibrium process. After removing the forward bias, the depletion region attempts to extend due to the Schottky contact. Consequently, more neutral trapping centers are ionized in the depletion layer and a large number of electrons are generated. Those electrons conversely repel the extension of the depletion layer. Finally, the extension of the depletion layer will be in dynamic balance with the repulsion by generated electrons. This is the possible reason why the LRS needs 72 hours to get stable as observed in Fig. 2. A direct evidence is the time dependent capacitance as shown in the inset of Fig. 4 (measured at 1 kHz with zero bias). The capacitance of the LRS decreased at the beginning, and became stable after 72 hours, which corresponds to the resistance change of the LRS in Fig. 2. The capacitance of the HRS is much more stable, resulting in the stable resistance of the HRS in Fig. 2. Because the capacitance change is due to the modification of the depletion layer width, the nonvolatility of the resistive switching is directly related to the dynamic balance of the depletion layer.

It has to be pointed out that BFO is ferroelectric. Therefore, we could not completely exclude the effect of polarization on the observed resistive switching. Choi *et al*. reported that the direction of bulk electric polarization can influence the diode behavior of a single crystal BFO. However, in a polycrystalline BFO thin film, the relationship between the polarization and the transport property is difficult to be investigated due to the large number of grain boundaries, which may dominate the transport.[23] We also measured the polarization-electric (PE) loops of our BFO thin films (not shown). The coercive field is rather large due to the leakage current, and the voltage used to measure the I-V curve is not large enough to switch the polarization sufficiently. Therefore, the effect of ferroelectric polarization does not dominate the resistive switching in our polycrystalline BFO thin films.

IV. CONCLUSIONS

In summary, the forming-free bipolar resistance switching behavior in the Au/BFO/Pt stack has been demonstrated using I-V and pulsed voltage measurements. The different polarities of the external voltage induce an electron trapping or detrapping process, and consequently change the depletion width below the Au Schottky contact, which is revealed by the capacitance-voltage measurements and by time dependent capacitance measurements at zero bias. The present work can help to further understand the physical origin of bipolar switching in $BiFeO_3$ and in other thin film oxides with electron trapping centers.

**ACKNOWLEDGEMENTS**

Yao Shuai would like to thank the China Scholarship Council (grant number: 2009607011). Shengqiang Zhou, Danilo Bürger, and Heidemarie Schmidt thank the financial support from the Bundesministerium für Bildung und Forschung (BMBF grant number: 13N10144). We acknowledge fruitful discussions with Dr. Stefan Slesazeck and Prof. Thomas Mikolajick.

**FIGURE CAPTIONS:**

Fig. 1. Rectifying and hysteretic I-V curves of Au/BFO/Pt, Au is positively biased and Pt is grounded. The arrows indicate the sweep direction of the bias. The current is normalized with respect to the area of the Au top contact.

Fig. 2. Pulsed voltage measurement on the resistive switching behavior of the Au/BFO/Pt. Two nonvolatile resistance states are read out by a continuous reading pulse (+2V) in dependence on time. The current is normalized with respect to the area of the Au top contact. Inset: time dependence of the resistance ratio between HRS and LRS.

Fig. 3. I-V curves plotted on a (a) log (J/E)~log ($E^{1/2}$) scale in the positive bias range (PF conduction), and on a (b) log (J)~log ($E^{1/2}$) in negative bias range (SE conduction). The inset of (a) is log ($J/E^2$)~log (1/E) scale in positive bias range (FN conduction). The current is normalized with respect to the area of the Au top contact.

Fig. 4. C-V curves in two voltage sweeping directions. The inset shows the time-dependence of the capacitance of the LRS and HRS, respectively, at zero bias.

Tab. 1. PLD growth parameters of BFO thin films

| | |
|---|---|
| Temperature | 700℃ |
| Pressure | 13 mTorr |
| Laser fluence | 2 J/cm$^2$ |
| Repetition rate | 10 Hz |
| Target-substrate distance | 60 mm |

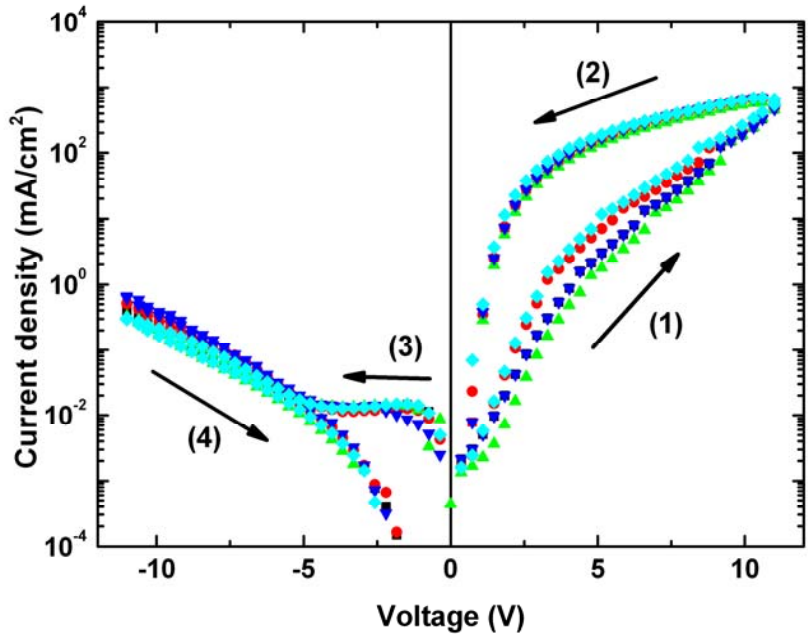

Fig. 1

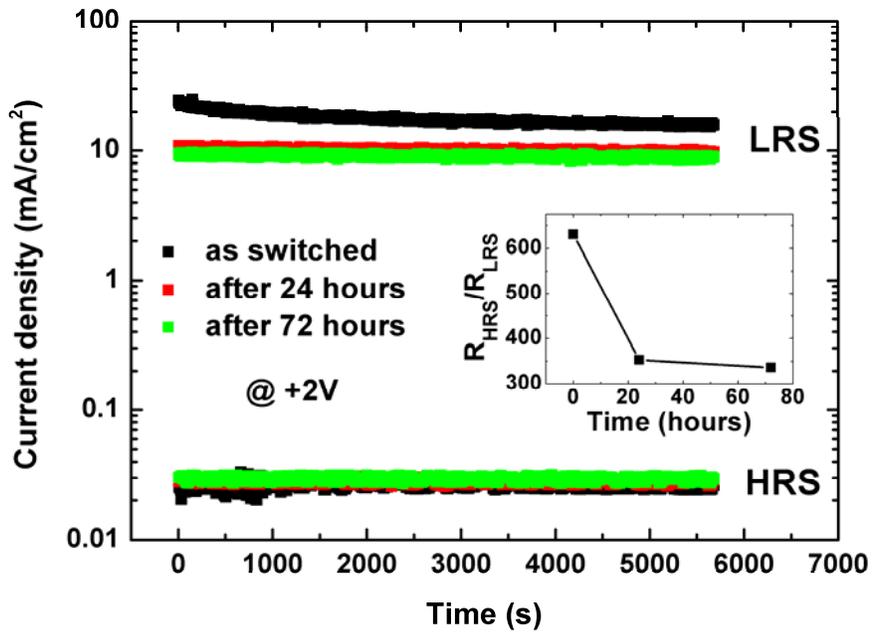

Fig. 2

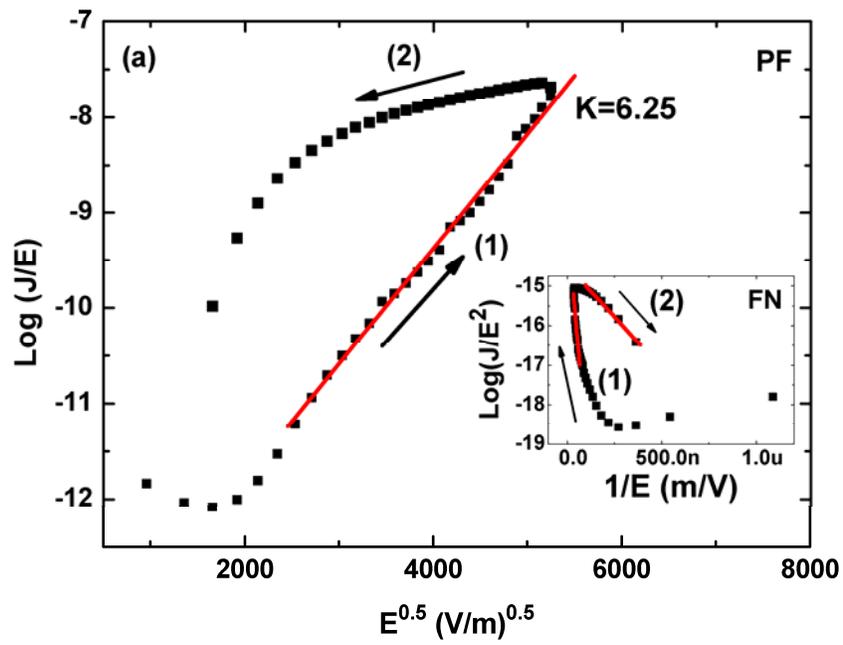

Fig. 3(a)

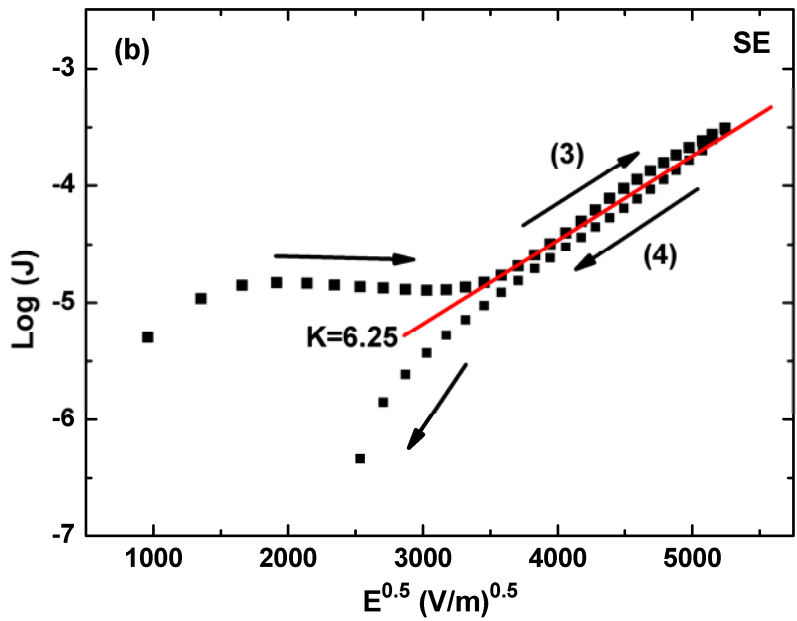

Fig. 3(b)

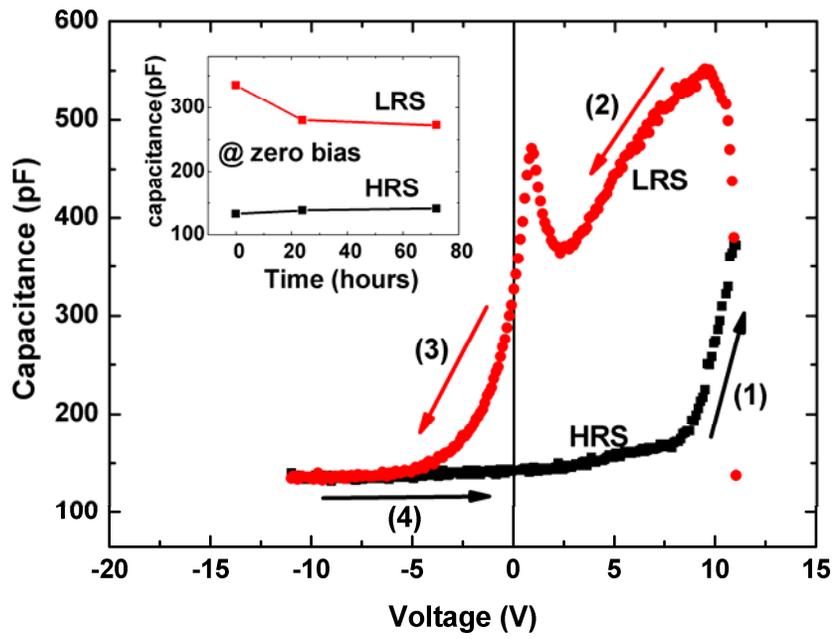

Fig. 4

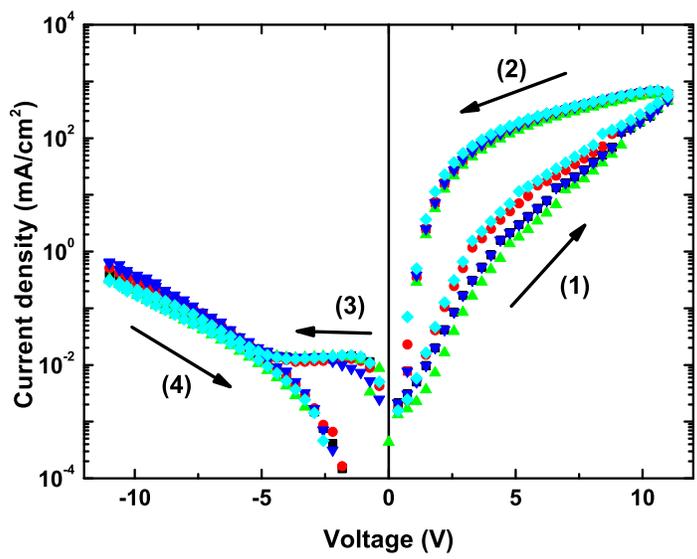

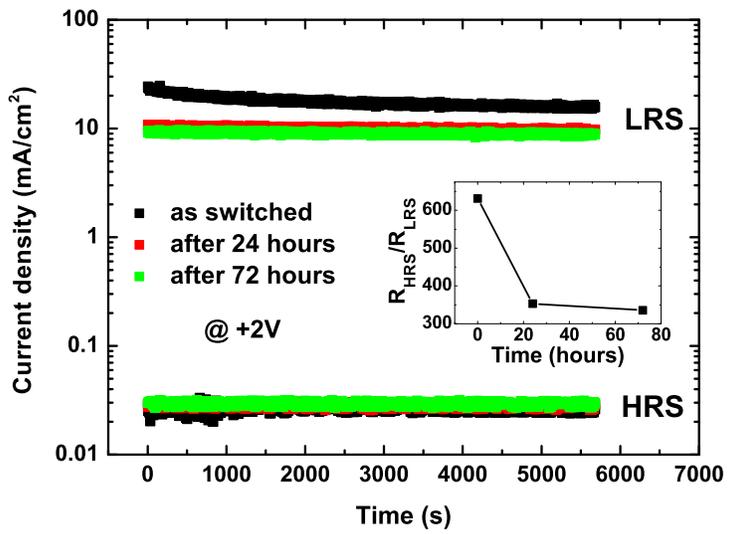

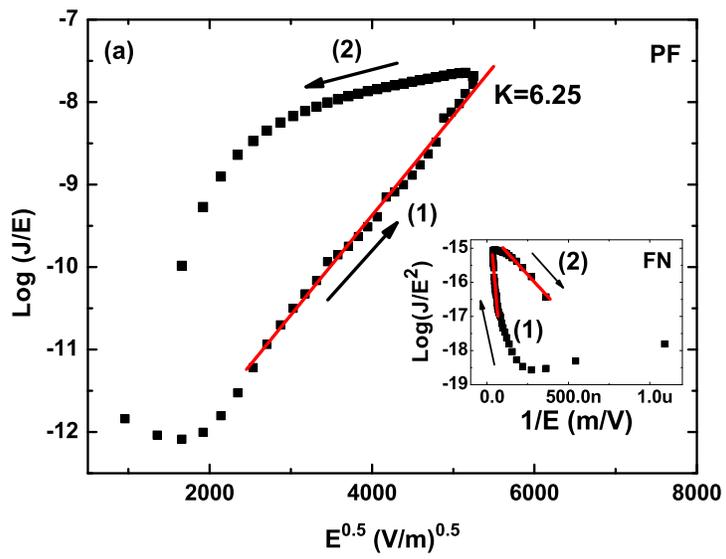

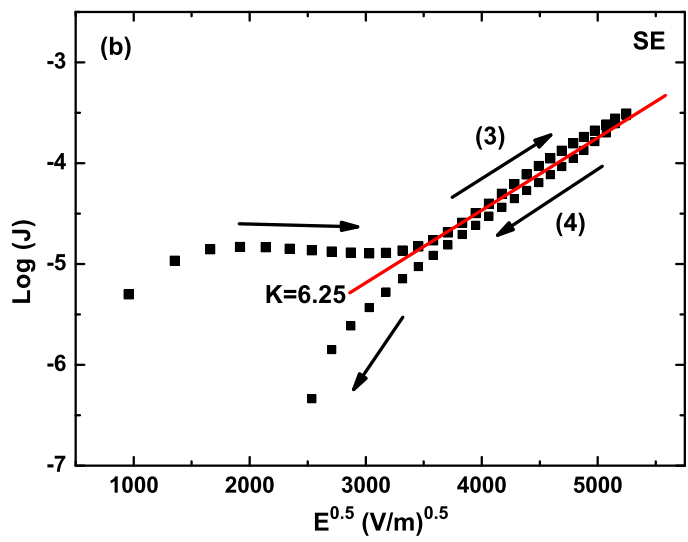

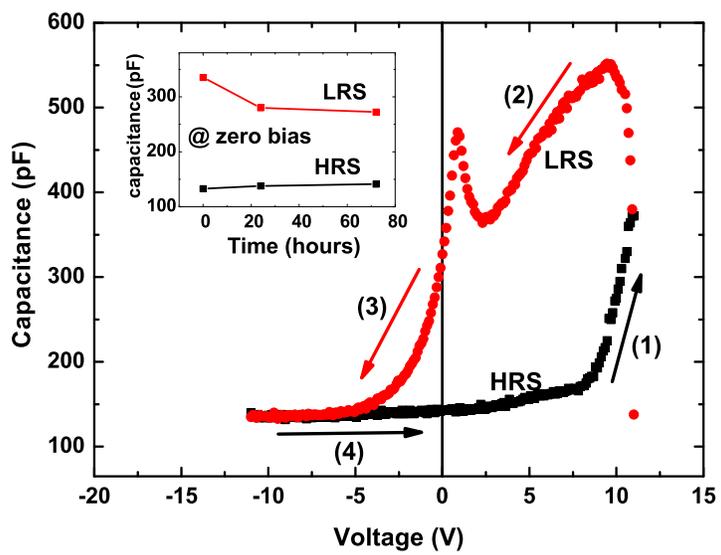